\documentclass[review,3p, 12pt]{elsarticle}

\usepackage{amssymb}

\biboptions{sort&compress}

\usepackage[bookmarks=false, colorlinks=true]{hyperref}
\usepackage{breakurl}

\usepackage[usenames,dvipsnames]{xcolor}
\usepackage[margin=10pt,font=small,labelfont=bf, indention=0.12cm, skip=5pt]{caption}
\newcommand\fnote[1]{\captionsetup{font=small}\caption*{#1}}
\usepackage{booktabs}
\usepackage[flushmargin]{footmisc}

\usepackage{dcolumn}

\journal{Physica A}

\begin{document}

\begin{frontmatter}


\title{Do wealth distributions follow power laws? \\Evidence from ``rich lists''}


\author{Michal Brzezinski}

\address{Faculty of Economic Sciences, University of Warsaw, Dluga 44/50, 00-241, Warsaw, Poland \ead{mbrzezinski@wne.uw.edu.pl}
\\
\vspace{1cm}
\begin{normalsize}
\end{normalsize}}

\begin{abstract}

We use data on wealth of the richest persons taken from the ``rich lists'' provided by business magazines like \textit{Forbes} to verify if upper tails of wealth distributions follow, as often claimed, a power-law behaviour. The data sets used cover the world's richest persons over 1996--2012, the richest Americans over 1988--2012, the richest Chinese over 2006--2012 and the richest Russians over 2004--2011. Using a recently introduced comprehensive empirical methodology for detecting power laws, which allows for testing goodness of fit as well as for comparing the power-law model with rival distributions, we find that a power-law model is consistent with data only in $35\%$ of the analysed data sets.  Moreover, even if wealth data are consistent with the power-law model, usually they are also consistent with some rivals like the log-normal or stretched exponential distributions.
\end{abstract}


\end{frontmatter}


\bigskip
\section{Introduction}\label{sec:intro}

The search for universal regularities in income and wealth distributions has started over one hundred years ago with the famous work of \citet{pareto1897cours}. His work suggested that the upper tails of income and wealth distributions follow a power law, which for a quantity $x$ is defined as a probability distribution $p(x)$ proportional to $x^{-\alpha}$ , with $\alpha>0$ being a positive shape parameter known as the Pareto (or power-law) exponent. Pareto's claim has been extensively tested empirically as well as studied theoretically \citep{chatterjee2005econophysics, chakrabarti2006econophysics, yakovenko2009colloquium, yakovenko2009econophysics, chakrabarti2013econophysics}. The emerging consensus in the empirical econophysics literature is that the bulk of income and wealth distributions seems to follow the log-normal or gamma distributions, while the upper tail follows power-law distribution. Recent empirical studies found a power-law behavior in the distribution of income in Australia \citep{Clementi200649, Banerjee200654}, Germany \citep{clementi2005a}, India \citep{Sinha2006555}, Italy \citep{clementi2005a, clementi2005, Clementi200649}, Japan \citep{souma2001universal, Aoyama2003352}, the UK \citep{dragulescu2001exponential, clementi2005a, richmond2006review}, and the USA \citep{dragulescu2001exponential, clementi2005a, silva2005temporal}. 
Another group of studies discovered a power-law structure of the upper tail of modern wealth distributions in China \citep{ning2007power}, France \citep{levy1998}, India \citep{Sinha2006555, Jayadev2008270}, Sweden \citep{Levy200342}, the UK \citep{levy1998, dragulescu2001exponential, Levy200342, coelho2005family}, and the USA \citep{levy1997new, levy1998, Levy200342, klass2007forbes}.  Surprisingly, analogous result were obtained for wealth distribution of aristocratic families in medieval Hungary \citep{Hegyi2007271} and for the distribution of house areas in ancient Egypt \citep{abul2002wealth}. 

However, as shown recently by \citet{clauset2009power} detecting power laws in empirical data may be a difficult task. Most of the existing empirical studies exploit the fact that the power-law distribution follows a straight line on a log-log plot with the power-law exponent equal to the absolute slope of the fitted line. The existence of power-law behaviour is often confirmed visually using such a plot, while the exponent is estimated using linear regression. Such approach suffers, however, from several drawbacks \citep{goldstein2004problems, clauset2009power}. First, the estimates of the slope
of the regression line may be very biased. Second, the standard $R^2$ statistic for the fitted regression line cannot be treated as a reliable goodness of fit test for the power-law behaviour. Third, even if traditional methods succeed in verifying that a power-law model is a good fit to a given data set, it is still possible that some alternative model fits the data better. A complete empirical analysis would therefore require conducting a     statistical comparison of the power-law model with  some other candidate distributions. 

Using a more refined methodology for measuring  power-law behaviour, \citet{clauset2009power} have shown that the distribution of wealth among the richest Americans in 2003 as compiled in \textit{Forbes}' annual US ``rich list'' is not fitted well by a power-law model.
Recently, \citet{ogwang2012wealth} has tested formally for a power-law behaviour in Forbes' data on the  wealth of the world's billionaires for the years from 2000 to 2009. He has found that the Kolmogorov-Smirnov, Anderson-Darling and $\chi^2$ goodness of fit tests all reject power-law behaviour for each of the data sets he used. However, \citet{ogwang2012wealth} has tested if the whole range of observations in his data sets follow a power-law behaviour, while in fact this may apply only to a subset of the very largest observations. A more appropriate methodology for detecting a power-law distribution would have, therefore, include a procedure for estimating a lower bound on the power-law behaviour.

The present paper uses a complete empirical methodology for detecting power laws introduced by \citet{clauset2009power} to verify if upper tails of wealth distributions obey the power-law model or if some alternative model fits the data better. We estimate both the power-law exponent and the lower bound on the power-law behaviour. We also use goodness of fit tests and compare power-law fits with fits of alternative models. We analyze a large number of data sets on wealth distributions published annually by \textit{Forbes} and other business magazines concerning wealth of 1) the richest persons in the world, 2) the richest Americans, 3) the richest Chinese, and 4) the richest Russians.  






The paper is organized as follows. Section \ref{sec2:methods} presents the statistical framework used for measuring and analyzing power-law behavior in empirical data introduced by \citet{clauset2009power}. Section \ref{section3:data} shortly describes our data sets drawn from the lists of the richest persons published by\textit{ Forbes} and other sources, while Section \ref{sec:results} provides the empirical analysis. Section \ref{conc} concludes.






\section{Statistical methods}
\label{sec2:methods}
 
In order to detect a power-law behaviour in wealth distributions we use a toolbox proposed by \citet{clauset2009power}. A density of continuous  power-law model is given by
\begin{equation}
p(x)=\frac{\alpha-1}{x_{min}}\left(\frac{x}{x_{min}}\right)^{-\alpha}.
\end{equation} 
 The maximum likelihood estimator (MLE) of the power-law exponent, $\alpha$,   is
\begin{equation}\hat
\alpha=1+n\left[ \sum_{i=1}^n\ln\frac{x_i}{x_{min}} \right],
\end{equation}
where $x_i, i=1, \dots , n $ are independent observations such that $x_i\geqslant x_{min}$.
The lower bound on the power-law behaviour, $x_{min}$, will be estimated using the following procedure. For each $x_i\geqslant x_{min}$, we estimate the exponent using the MLE and then we compute the well-known Kolmogorov-Smirnov (KS) statistic for the data and the fitted model. The estimate $\hat{x}_{min}$ is then chosen as a value of $x_i$ for which the KS statistic is the smallest.\footnote{The Kolmogorov-Smirnov statistic was also proposed by \citet{goldstein2004problems} as a goodness of fit test for the discrete power-law model assuming, however, that the lower bound on power-law behaviour is known.} The standard errors for estimated parameters are computed with standard bootstrap methods with 10,000 replications. 

The next step in measuring power laws involves testing goodness of fit. A positive result of such a test allows to conclude that a power-law model is consistent with a given data set. Following \citet{clauset2009power} again, we use a test based on a semi-parametric bootstrap approach. The procedure starts with fitting a power-law model to data using the MLE for $\alpha$ and the KS-based estimator for $x_{min}$ and calculating a KS statistic for this fit, $ks$. Next, a large number of bootstrap data sets is generated that follow the originally fitted power-law model above the estimated $x_{min}$ and have the same non-power-law distribution as the original data set below $\hat x_{min}$. Then, power-law models are fitted to each of the generated data sets using the same methods as for the original data set and the KS statistics are calculated.
The fraction of data sets for which their own KS statistic is larger than $ks$ is  the $p$-value of the test. The power-law hypothesis is rejected if this $p$-value is smaller than some chosen threshold. Following \citet{clauset2009power}, we rule out the power-law model if the estimated $p$-value for this test is smaller than 0.1. In our computations, we use 4,999 generated  data sets. 

If  the goodness of fit test rejects the power-law hypothesis, we may conclude that the power law has not been found. However, if  a data set is well fit by a power law, the question remains if there is other alternative distribution, which is equally good or better fit to this data set. We need, therefore, to fit some rival distributions and evaluate which distribution gives a better fit. To this end, \citet{clauset2009power} use the likelihood ratio test proposed by \citet{vuong1989likelihood}. The test computes the logarithm of the ratio of the likelihoods of the data under two competing distributions, $LR$, which is negative or positive depending on which model fits data better. \citet{vuong1989likelihood} showed that in the case of non-nested models the normalized log-likelihood ratio $NLR =n^{-1/2}LR/\sigma$, where $\sigma$ is the estimated standard deviation of $R$, has a limit standard normal distribution. This result can be used to compute a $p$-value for the test discriminating between the competing models. In case of nested models, \citet{vuong1989likelihood} shows that $2LR$ has a limit a chi-squared distribution.

Each of the estimators and tests described above has been tested with good effects by \citet{clauset2009power} using Monte Carlo simulations.\footnote{The Stata software implementing all methods described in this section is available from the author upon request. The original power-law-testing Matlab and R software written by Aaron Clauset and Cosma R. Shalizi can be obtained from \href{http://tuvalu.santafe.edu/~aaronc/powerlaws/}{http://tuvalu.santafe.edu/\textasciitilde aaronc/powerlaws/}.}  

\section{Wealth data from the ``rich lists''}
\label{section3:data}

In several countries business magazines publish annual lists of the richest individuals. The oldest and the most famous one is the Forbes 400 Richest Americans list, which started in 1982. Other ``rich lists'' published by Forbes include   the World's Billionaires and the 400 Richest Chinese. These lists provide rankings of rich individuals according to their net worth defined as a sum of their assets minus their debts.
We use annual data from the Forbes 400 Richest Americans list for the period 1988-2012, from the Forbes World's Billionaires list for the period 1996--2012 and from the Forbes 400 Richest Chinese list for 2006--2012. 
In addition, we use 2004--2011 data from the list of top Russian billionaires published by the Russian magazine Finans (www.finansmag.ru). Descriptive statistics for our data sets are presented using beanplots \citep{kampstra2008} in Appendix A.

\section{Results}
\label{sec:results}

\begin{figure}[h]
\centering
\caption{Power-law exponents and goodness of fit tests for wealth data\label{results}$^{a}$}
\includegraphics[scale=1]{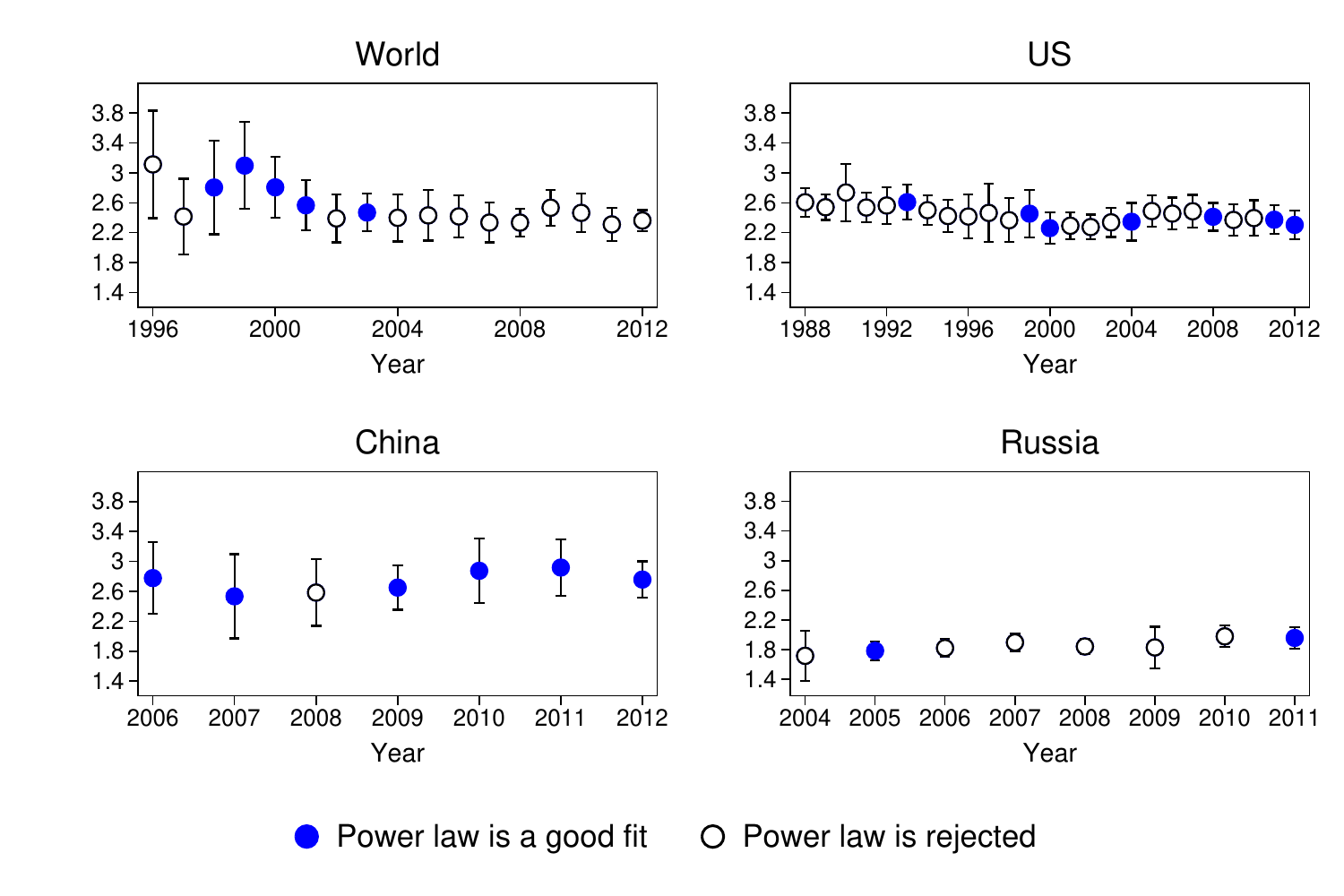}
\fnote{$^{a}$ Vertical bars show 95\% confidence intervals.}
\end{figure}

\begin{figure}[h]
\centering
\caption{The complementary cumulative distribution functions and their power-law fits\label{figure2}}
\includegraphics[scale=1]{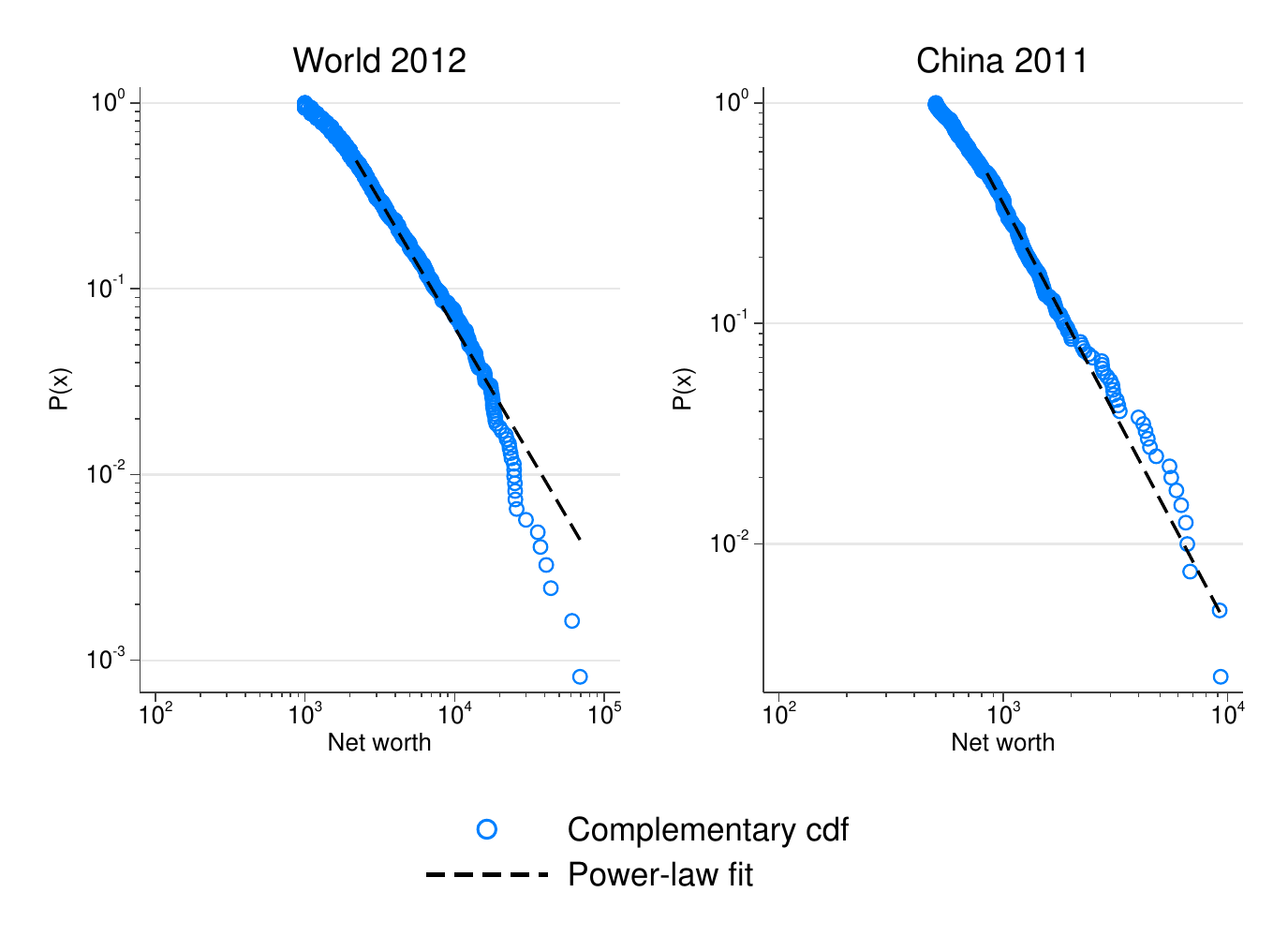}
\end{figure}

Power-law fits to our data sets are shown in Figure \ref{results}.
The values of the power-law exponent are rather stable over time for all four groups of data sets studied. However, except for Russia, the estimated exponents are substantially higher than usually found in the previous  literature on power-law behaviour of wealth distributions.\footnote{\citet{richmond2006review} found that the estimated values of the power-law exponent range from 0.5 to 1.5 for wealth distributions and from about 1.5 to 3 for income distributions.} 
In particular, the average estimate of the exponent for the world's richest persons is $2.5$, while the averages for the US, China and Russia are, respectively, $2.4$, $2.7$ and $1.9$. This result is a consequence of the fact that previous studies have rarely attempted to estimate $x_{min}$ and instead often fitted power-law models to all available observations. However, estimating $x_{min}$ using KS-based approach as described in Section \ref{sec2:methods} leads to a substantially smaller range of observations that may potentially follow the power-law behaviour. According to our estimates, on average only about  306 observations (44\% of all available observations)  are above $\hat{x}_{min}$ in case of data sets covering the world's richest persons. The average number of observations above $\hat{x}_{min}$ is 268 ($60\%$), 261 ($57\%$) and 220 ($55\%$) for the US, Russia and China, respectively.
The most striking conclusion from Figure \ref{results} is that the majority of the data sets for the world's richest persons, the richest Americans, and the richest Russians are not fitted well by the power-law model according to the goodness of fit test used. For Russia only $25\%$ of data sets seem to follow a power-law behaviour, while for the richest persons in the world and for the richest Americans the number is in the range from $28$ to $29\%$. Only for China in all but one case wealth distribution seems to follow a power-law model, but the period under study for this country is the shortest. These results suggest that, at least for the data sets drawn from the ``rich lists'', wealth distributions often do not follow the power-law model and that testing goodness of fit should always precede a declaration that a power-law behaviour of a wealth distribution was found. Figure \ref{figure2} shows typical examples when a power law is not a good fit for our data sets (left panel, goodness of fit test $p$-value = 0.02) and for the case when is seems to be a good fit (right panel, $p$-value = 0.64).

Our results are inconsistent with those of \citet{ogwang2012wealth}, who found that the power-law behaviour of the wealth of the world's richest persons 
according to Forbes' data in every year between 2000 and 2009 is ruled out by conventional goodness of fit tests. On the contrary, we find that the wealth of the world's billionaires is fitted well by a power-law model in 2000, 2001 and 2003. This inconsistency can be explained by noticing that \citet{ogwang2012wealth} has not estimated the lower bound on the power-law behaviour, $x_{min}$, but fitted power-law models to the whole range of Forbes' observations. However, fixing $x_{min}$ at the minimum wealth level in Forbes' data seems to be statistically unjustified.  

The results of the likelihood ratio tests for all 20 data sets that passed the goodness of fit test (with $p$-value $>0.1$) are given in Table \ref{table1}. We have followed \citet{clauset2009power} in choosing the following alternative distributions: log-normal, exponential, stretched exponential and power-law with exponential cut-off.\footnote{See \citet{clauset2009power} for definitions of these distributions. The pure power-law model is a nested within the power-law with exponential cut-off model and for this reason the latter always provides a fit at least as good as the former. The $LR$ statistic for comparing these models will therefore be always negative or zero.} 
Positive (negative) values of $NLR$ or $LR$ mean that the power-law model gives a better (worse) fit compared to a given alternative. If the $p$-value for the likelihood ratio test is small, then we may reject the model which gives a worse fit to data. If the $p$-value is larger than the chosen level, which is set to $0.1$ in our analysis, then we are not able to choose between the compared models.  

\begin{table}[htbp] \footnotesize
\caption{\label{table1} Power-law vs. other models for the upper tail of wealth distributions$^a$ }
\centering 
\begin{tabular}{lcccccccccc}
\toprule
\multicolumn{1}{l}{Data set}  & \multicolumn{1}{c}{Power law} & \multicolumn{2}{c}{Log-normal} & \multicolumn{2}{c}{Exponential} & 
\multicolumn{2}{c}{Stretched} & \multicolumn{2}{c}{Power law} & 
\multicolumn{1}{c}{Support for} \\
&&&&&& \multicolumn{2}{c}{exponential} & \multicolumn{2}{c}{with cut-off} &
\multicolumn{1}{c}{power law}\\
 \cmidrule(r){2-2} \cmidrule(r){3-4}  \cmidrule(r){5-6}  \cmidrule(r){7-8}
  \cmidrule(r){9-10}  
& \multicolumn{1}{c}{$p$}  &\multicolumn{1}{c}{$NLR$} &\multicolumn{1}{c}{$p$}
&\multicolumn{1}{c}{$NLR$} &\multicolumn{1}{c}{$p$} &\multicolumn{1}{c}{$NLR$} &\multicolumn{1}{c}{$p$}
&\multicolumn{1}{c}{$LR$} &\multicolumn{1}{c}{$p$} & \\
\midrule
\multicolumn{1}{l}{\textit{World}} \\
1998 & $0.981$ & $0.026$ & $0.979$ & $2.043$ & $0.041$ & $-0.043$ & $0.966$ & $-0.068$ & $0.713$ &moderate\\
1999 & $0.977$ & $0.483$ & $0.629$ & $1.447$ & $0.148$ & $0.146$ & $0.884$ & $0.000$ & $1.000$ & moderate \\
2000 &$0.824$&$-0.144$&$0.886$&$2.313$&$0.021$&$-0.158$&$0.874$&$-0.141$&$0.595$& moderate\\
2001 &$0.490$&$-0.544$&$0.586$&$2.587$&$0.010$&$-0.554$&$0.580$&$-0.777$&$0.212$& moderate\\
2003 &$0.154$&$-0.929$&$0.353$&$2.623$&$0.009$&$-0.955$&$0.340$&$-1.715$&$0.064$& with cut-off\\
\multicolumn{1}{l}{\textit{US}} \\
1993 &$0.297$&$-0.680$&$0.496$&$3.146$&$0.002$&$-0.694$&$0.488$&$-0.957$&$0.167$& moderate\\
1999 &$0.506$&$-0.116$&$0.908$&$2.325$&$0.020$&$-0.137$&$0.891$&$-0.141$&$0.595$& moderate\\
2000 &$0.189$&$-0.361$&$0.718$&$3.630$&$0.000$&$-0.361$&$0.718$&$-0.660$&$0.251$& moderate\\
2004 &$0.315$&$-0.431$&$0.666$&$2.429$&$0.015$&$-0.441$&$0.659$&$-0.637$&$0.259$& moderate\\
2008 &$0.268$&$-0.879$&$0.379$&$2.171$&$0.030$&$-0.902$&$0.367$&$-1.533$&$0.080$& with cut-off\\
2011 &$0.381$&$-0.307$&$0.759$&$4.615$&$0.000$&$-0.316$&$0.752$&$-0.721$&$0.230$& moderate\\
2012 &$0.386$&$-0.791$&$0.429$&$2.831$&$0.005$&$-0.813$&$0.416$&$-1.432$&$0.091$& with cut-off\\
  \multicolumn{1}{l}{\textit{China}} \\
2006 & $0.377$ & $-0.817$ & $0.414$ &$1.561$ & $0.119$ & $-0.850$ & $0.396$ & $-1.090$ & $0.140$ & moderate\\
2007 &$0.244$&$-0.699$&$0.484$&$1.394$&$0.163$&$-0.705$&$0.481$&$-0.800$&$0.206$& moderate\\
2009 &$0.295$&$-0.940$&$0.347$&$2.883$&$0.004$&$-0.975$&$0.330$&$-1.658$&$0.069$& with cut-off\\
2010 &$0.168$&$-0.469$&$0.639$&$2.739$&$0.006$&$-0.479$&$0.632$&$-0.656$&$0.252$& moderate\\
2011 &$0.636$&$0.365$&$0.715$&$3.820$&$0.000$&$0.084$&$0.933$&$-0.119$&$0.626$& moderate\\
2012 &$0.676$&$-0.407$&$0.684$&$3.633$&$0.000$&$-0.412$&$0.681$&$-0.497$&$0.319$& moderate\\
\multicolumn{1}{l}{\textit{Russia}} \\
2005 &$0.101$&$-1.596$&$0.110$&$3.200$&$0.001$&$-1.664$&$0.096$&$nc$&$-$& moderate\\
2011 &$0.661$&$-0.860$&$0.390$&$6.270$&$0.000$&$-0.886$&$0.375$&$nc$&$-$& moderate\\
\bottomrule
\end{tabular}
\medskip
\\
\parbox{16.5cm}{$^{a}$ \scriptsize The first column gives $p$-values for the goodness of fit test for the power-law behaviour. For each non-nested alternative distribution we give the normalized log-likelihood ratio ($NLR$), while for the power law with exponential cut-off we give the log-likelihood ratio ($LR$). ``nc'' denotes a non-convergence of the MLE for a distribution. We also present $p$-values ($p$) for the significance of $NLR$ or $LR$. The last column presents the final judgement using the terminology of \citet{clauset2009power}: ``moderate'' means that power law is a good fit but so are some plausible alternatives; ``with cut-off'' means that the power law with exponential cut-off is favoured over the pure power law, but there are also other plausible models. None of our data sets can be labelled ``good'', which means that that the power law is a good fit and that none of the alternatives considered is plausible.}
\end{table}

Results from Table \ref{table1} show there is no data set for which we may conclude that it is fitted well by a power-law model and that it is there is no plausible alternative model for it. There are only two data sets (the world's richest persons in 1999 and the richest Chinese in 2011) for which the sign of the $NLR$ statistic suggests that the power law is better than the the log-normal, exponential and stretched exponential distributions, but $p$-values for the statistic are in most cases so large that the tests are inconclusive. More generally, the exponential distribution can be ruled out as a plausible model for our wealth data in all but three cases (the world's richest persons in 1999 and the richest Chinese in 2006 and 2007). On the other hand, the log-normal distribution appears to be empirically indistinguishable from the power law in our data -- the $p$-values for the relevant tests are always larger than $0.11$. Similar conclusion can be drawn for a comparison between the stretched exponential and the power law with the possible exception of data for Russia in 2005, for which the stretched exponential distribution seems to be marginally favored over the power law.
The power law with an exponential cut-off seems to fit the data better than the pure power law in four cases (the world's richest persons in 2003, the richest Americans in 2008 and 2012 and the richest Chinese in 2009). However, in these four cases the log-normal and stretched exponential distributions are also plausible fits to data.

In overall, none of the 57 data sets on wealth distribution analysed in this paper can be reliably described as fitted best by a power-law model. Only about $35\%$ of data sets can be plausible considered to follow a power-law distribution, but even among these data sets the power law is empirically indistinguishable from the log-normal and stretched exponential distributions.

These results suggest that the hypothesis that upper tails of wealth distributions, at least when measured using data from ``rich lists'', follow a power-law behavior is statistically doubtful. It seems obvious that this hypothesis should no longer be assumed before conducting an empirical analysis of a given data set using tools similar to those of \citet{clauset2009power}. The existence of popular software implementing such empirical methods should    make this task easier. The results of this paper seem also to cast some doubt on the theoretical literature
in  econophysics and economics that provides a theoretical structure for power-law behaviour of top wealth values. Theoretical models that make room also for some other distributions (especially the log-normal and stretched exponential distributions) describing upper tail of wealth distribution  may be equally worth consideration.
\section{Conclusions}
\label{conc}

In this paper, we have used a large number of data sets on wealth distribution taken from the lists of the richest persons published annually by business magazines like Forbes. Using recently developed empirical methodology for detecting power-law behaviour introduced by \citet{clauset2009power}, we have found that top wealth values follow the power-law behaviour only in $35\%$ of analysed cases. Moreover, even if the data do not rule out the power-law model usually the evidence in its favour is not conclusive -- some rivals, most notably the log-normal and stretched exponential distributions, are also plausible fits to wealth data.



\section*{Acknowledgements}
\label{Acknowledgements}
\noindent
I would like to acknowledge gratefully the Matlab and R software written by Aaron Clauset and Cosma R. Shalizi, which implements methods described in Section \ref{sec2:methods}. The software can be obtained from \href{http://tuvalu.santafe.edu/~aaronc/powerlaws/}{http://tuvalu.santafe.edu/\textasciitilde aaronc/powerlaws/}. I also thank Moshe Levy for sharing data from Forbes 400 Richest Americans lists. I also appreciate the helpful comments of the participants of the 
32nd International Association for Research in Income and Wealth conference, Boston, USA, August 5-11, 2012. All remaining errors are my own. This work was supported by Polish National Science Centre grant no. 2011/01/B/HS4/02809.



\section*{References}

\bibliographystyle{model1-num-names}
\bibliography{PowerLawWealth}

\begin{thebibliography}{30}
\expandafter\ifx\csname natexlab\endcsname\relax\def\natexlab#1{#1}\fi
\providecommand{\bibinfo}[2]{#2}
\ifx\xfnm\relax \def\xfnm[#1]{\unskip,\space#1}\fi
\bibitem[{Pareto(1897)}]{pareto1897cours}
\bibinfo{author}{V.~Pareto}, \bibinfo{title}{{Cours d'Economie Politique}},
  \bibinfo{publisher}{F. Rouge, Lausanne}, \bibinfo{year}{1897}.
\bibitem[{Chatterjee et~al.(2005)Chatterjee, Yarlagadda, and
  Chakrabarti}]{chatterjee2005econophysics}
\bibinfo{editor}{A.~Chatterjee}, \bibinfo{editor}{S.~Yarlagadda},
  \bibinfo{editor}{B.~Chakrabarti} (Eds.), \bibinfo{title}{{Econophysics of
  wealth distributions}}, \bibinfo{publisher}{Springer Verlag, Milan},
  \bibinfo{year}{2005}.
\bibitem[{Chakrabarti et~al.(2006)Chakrabarti, Chakraborti, and
  Chatterjee}]{chakrabarti2006econophysics}
\bibinfo{editor}{B.~Chakrabarti}, \bibinfo{editor}{A.~Chakraborti},
  \bibinfo{editor}{A.~Chatterjee} (Eds.), \bibinfo{title}{{Econophysics and
  sociophysics: trends and perspectives}}, \bibinfo{publisher}{Wiley-VCH,
  Berlin}, \bibinfo{year}{2006}.
\bibitem[{Yakovenko and Rosser~Jr(2009)}]{yakovenko2009colloquium}
\bibinfo{author}{V.~Yakovenko}, \bibinfo{author}{J.~Rosser~Jr},
\newblock \bibinfo{title}{{Colloquium: Statistical mechanics of money, wealth,
  and income}},
\newblock \bibinfo{journal}{Reviews of Modern Physics} \bibinfo{volume}{81}
  (\bibinfo{year}{2009}) \bibinfo{pages}{1703--1725}.
\bibitem[{Yakovenko(2009)}]{yakovenko2009econophysics}
\bibinfo{author}{V.~M. Yakovenko},
\newblock \bibinfo{title}{{Econophysics, statistical mechanics approach to}},
\newblock in: \bibinfo{editor}{R.~A. Meyers} (Ed.),
  \bibinfo{booktitle}{Encyclopedia of Complexity and Systems Science},
  \bibinfo{publisher}{Springer, Berlin}, \bibinfo{year}{2009}, pp.
  \bibinfo{pages}{2800--2826}.
\bibitem[{Chakrabarti et~al.(2013)Chakrabarti, Chakraborti, Chakravarty, and
  Chatterjee}]{chakrabarti2013econophysics}
\bibinfo{author}{B.~K. Chakrabarti}, \bibinfo{author}{A.~Chakraborti},
  \bibinfo{author}{S.~R. Chakravarty}, \bibinfo{author}{A.~Chatterjee},
  \bibinfo{title}{Econophysics of Income and Wealth Distributions},
  \bibinfo{publisher}{Cambridge University Press, Cambridge},
  \bibinfo{year}{2013}.
\bibitem[{Clementi et~al.(2006)Clementi, Matteo, and
  Gallegati}]{Clementi200649}
\bibinfo{author}{F.~Clementi}, \bibinfo{author}{T.~D. Matteo},
  \bibinfo{author}{M.~Gallegati},
\newblock \bibinfo{title}{The power-law tail exponent of income distributions},
\newblock \bibinfo{journal}{Physica A} \bibinfo{volume}{370}
  (\bibinfo{year}{2006}) \bibinfo{pages}{49--53}.
\bibitem[{Banerjee et~al.(2006)Banerjee, Yakovenko, and
  Matteo}]{Banerjee200654}
\bibinfo{author}{A.~Banerjee}, \bibinfo{author}{V.~M. Yakovenko},
  \bibinfo{author}{T.~D. Matteo},
\newblock \bibinfo{title}{A study of the personal income distribution in
  australia},
\newblock \bibinfo{journal}{Physica A} \bibinfo{volume}{370}
  (\bibinfo{year}{2006}) \bibinfo{pages}{54--59}.
\bibitem[{Clementi and Gallegati(2005)}]{clementi2005a}
\bibinfo{author}{F.~Clementi}, \bibinfo{author}{M.~Gallegati},
\newblock \bibinfo{title}{{Pareto's law of income distribution: Evidence for
  Germany, the United Kingdom, and the United States}},
\newblock in:  \cite{chatterjee2005econophysics}, pp. \bibinfo{pages}{3--14}.
\bibitem[{Sinha(2006)}]{Sinha2006555}
\bibinfo{author}{S.~Sinha},
\newblock \bibinfo{title}{Evidence for power-law tail of the wealth
  distribution in india},
\newblock \bibinfo{journal}{Physica A} \bibinfo{volume}{359}
  (\bibinfo{year}{2006}) \bibinfo{pages}{555--562}.
\bibitem[{Clementi and Gallegati(2005)}]{clementi2005}
\bibinfo{author}{F.~Clementi}, \bibinfo{author}{M.~Gallegati},
\newblock \bibinfo{title}{Power law tails in the italian personal income
  distribution},
\newblock \bibinfo{journal}{Physica A} \bibinfo{volume}{350}
  (\bibinfo{year}{2005}) \bibinfo{pages}{427--438}.
\bibitem[{Souma(2001)}]{souma2001universal}
\bibinfo{author}{W.~Souma},
\newblock \bibinfo{title}{{Universal structure of the personal income
  distribution}},
\newblock \bibinfo{journal}{Fractals} \bibinfo{volume}{9}
  (\bibinfo{year}{2001}) \bibinfo{pages}{463--470}.
\bibitem[{Aoyama et~al.(2003)Aoyama, Souma, and Fujiwara}]{Aoyama2003352}
\bibinfo{author}{H.~Aoyama}, \bibinfo{author}{W.~Souma},
  \bibinfo{author}{Y.~Fujiwara},
\newblock \bibinfo{title}{Growth and fluctuations of personal and company's
  income},
\newblock \bibinfo{journal}{Physica A} \bibinfo{volume}{324}
  (\bibinfo{year}{2003}) \bibinfo{pages}{352--358}. \bibinfo{note}{Proceedings
  of the International Econophysics Conference}.
\bibitem[{Dr\v{a}gulescu and Yakovenko(2001)}]{dragulescu2001exponential}
\bibinfo{author}{A.~Dr\v{a}gulescu}, \bibinfo{author}{V.~Yakovenko},
\newblock \bibinfo{title}{{Exponential and power-law probability distributions
  of wealth and income in the United Kingdom and the United States}},
\newblock \bibinfo{journal}{Physica A} \bibinfo{volume}{299}
  (\bibinfo{year}{2001}) \bibinfo{pages}{213--221}.
\bibitem[{Richmond et~al.(2006)Richmond, Hutzler, Coelho, and
  Repetowicz}]{richmond2006review}
\bibinfo{author}{P.~Richmond}, \bibinfo{author}{S.~Hutzler},
  \bibinfo{author}{R.~Coelho}, \bibinfo{author}{P.~Repetowicz},
\newblock \bibinfo{title}{{A review of empirical studies and models of income
  distributions in society}},
\newblock in:  \cite{chakrabarti2006econophysics}, pp.
  \bibinfo{pages}{131--160}.
\bibitem[{Silva and Yakovenko(2005)}]{silva2005temporal}
\bibinfo{author}{A.~Silva}, \bibinfo{author}{V.~Yakovenko},
\newblock \bibinfo{title}{{Temporal evolution of the" thermal" and"
  superthermal" income classes in the USA during 1983-2001}},
\newblock \bibinfo{journal}{Europhysics Letters} \bibinfo{volume}{69}
  (\bibinfo{year}{2005}) \bibinfo{pages}{304--310}.
\bibitem[{Ning and You-Gui(2007)}]{ning2007power}
\bibinfo{author}{D.~Ning}, \bibinfo{author}{W.~You-Gui},
\newblock \bibinfo{title}{{Power-law Tail in the Chinese Wealth Distribution}},
\newblock \bibinfo{journal}{Chinese Physics Letters} \bibinfo{volume}{24}
  (\bibinfo{year}{2007}) \bibinfo{pages}{2434--2436}.
\bibitem[{Levy(1998)}]{levy1998}
\bibinfo{author}{S.~Levy}, \bibinfo{title}{{Wealthy People and Fat Tails: An
  Explanation for the L\'{e}vy Distribution of Stock Returns}},
  \bibinfo{type}{University of California at Los Angeles, Anderson Graduate
  School of Management} \bibinfo{number}{1118}, Anderson Graduate School of
  Management, UCLA, \bibinfo{year}{1998}.
\bibitem[{Jayadev(2008)}]{Jayadev2008270}
\bibinfo{author}{A.~Jayadev},
\newblock \bibinfo{title}{{A power law tail in India's wealth distribution:
  Evidence from survey data}},
\newblock \bibinfo{journal}{Physica A} \bibinfo{volume}{387}
  (\bibinfo{year}{2008}) \bibinfo{pages}{270--276}.
\bibitem[{Levy(2003)}]{Levy200342}
\bibinfo{author}{M.~Levy},
\newblock \bibinfo{title}{Are rich people smarter?},
\newblock \bibinfo{journal}{Journal of Economic Theory} \bibinfo{volume}{110}
  (\bibinfo{year}{2003}) \bibinfo{pages}{42 -- 64}.
\bibitem[{Coelho et~al.(2005)Coelho, Neda, Ramasco, and
  Augustasantos}]{coelho2005family}
\bibinfo{author}{R.~Coelho}, \bibinfo{author}{Z.~Neda},
  \bibinfo{author}{J.~Ramasco}, \bibinfo{author}{M.~Augustasantos},
\newblock \bibinfo{title}{A family-network model for wealth distribution in
  societies},
\newblock \bibinfo{journal}{Physica A} \bibinfo{volume}{353}
  (\bibinfo{year}{2005}) \bibinfo{pages}{515--528}.
\bibitem[{Levy and Solomon(1997)}]{levy1997new}
\bibinfo{author}{M.~Levy}, \bibinfo{author}{S.~Solomon},
\newblock \bibinfo{title}{{New evidence for the power-law distribution of
  wealth}},
\newblock \bibinfo{journal}{Physica A} \bibinfo{volume}{242}
  (\bibinfo{year}{1997}) \bibinfo{pages}{90--94}.
\bibitem[{Klass et~al.(2007)Klass, Biham, Levy, Malcai, and
  Solomon}]{klass2007forbes}
\bibinfo{author}{O.~Klass}, \bibinfo{author}{O.~Biham},
  \bibinfo{author}{M.~Levy}, \bibinfo{author}{O.~Malcai},
  \bibinfo{author}{S.~Solomon},
\newblock \bibinfo{title}{{The Forbes 400, the Pareto power-law and efficient
  markets}},
\newblock \bibinfo{journal}{The European Physical Journal B}
  \bibinfo{volume}{55} (\bibinfo{year}{2007}) \bibinfo{pages}{143--147}.
\bibitem[{Hegyi et~al.(2007)Hegyi, Néda, and Santos}]{Hegyi2007271}
\bibinfo{author}{G.~Hegyi}, \bibinfo{author}{Z.~Néda}, \bibinfo{author}{M.~A.
  Santos},
\newblock \bibinfo{title}{{Wealth distribution and Pareto's law in the
  Hungarian medieval society}},
\newblock \bibinfo{journal}{Physica A} \bibinfo{volume}{380}
  (\bibinfo{year}{2007}) \bibinfo{pages}{271--277}.
\bibitem[{Abul-Magd(2002)}]{abul2002wealth}
\bibinfo{author}{A.~Abul-Magd},
\newblock \bibinfo{title}{{Wealth distribution in an ancient Egyptian
  society}},
\newblock \bibinfo{journal}{Physical Review E} \bibinfo{volume}{66}
  (\bibinfo{year}{2002}) \bibinfo{pages}{57104}.
\bibitem[{Clauset et~al.(2009)Clauset, Shalizi, and Newman}]{clauset2009power}
\bibinfo{author}{A.~Clauset}, \bibinfo{author}{C.~R. Shalizi},
  \bibinfo{author}{M.~E.~J. Newman},
\newblock \bibinfo{title}{Power-law distributions in empirical data},
\newblock \bibinfo{journal}{SIAM Review} \bibinfo{volume}{51}
  (\bibinfo{year}{2009}) \bibinfo{pages}{661--703}.
\bibitem[{Goldstein et~al.(2004)Goldstein, Morris, and
  Yen}]{goldstein2004problems}
\bibinfo{author}{M.~Goldstein}, \bibinfo{author}{S.~Morris},
  \bibinfo{author}{G.~Yen},
\newblock \bibinfo{title}{{Problems with fitting to the power-law
  distribution}},
\newblock \bibinfo{journal}{The European Physical Journal B}
  \bibinfo{volume}{41} (\bibinfo{year}{2004}) \bibinfo{pages}{255--258}.
\bibitem[{Ogwang(2013)}]{ogwang2012wealth}
\bibinfo{author}{T.~Ogwang},
\newblock \bibinfo{title}{{Is the wealth of the world's billionaires
  Paretian?}},
\newblock \bibinfo{journal}{Physica A} \bibinfo{volume}{392}
  (\bibinfo{year}{2013}) \bibinfo{pages}{757--762}.
\bibitem[{Vuong(1989)}]{vuong1989likelihood}
\bibinfo{author}{Q.~Vuong},
\newblock \bibinfo{title}{Likelihood ratio tests for model selection and
  non-nested hypotheses},
\newblock \bibinfo{journal}{Econometrica}  (\bibinfo{year}{1989})
  \bibinfo{pages}{307--333}.
\bibitem[{Kampstra(2008)}]{kampstra2008}
\bibinfo{author}{P.~Kampstra},
\newblock \bibinfo{title}{Beanplot: A boxplot alternative for visual comparison
  of distributions},
\newblock \bibinfo{journal}{Journal of Statistical Software}
  \bibinfo{volume}{28} (\bibinfo{year}{2008}).

\end{thebibliography}






\newpage

\appendix
\setcounter{figure}{0} \renewcommand{\thefigure}{A.\arabic{figure}}
\label{app}

\section{Descriptive statistics for wealth data from the ``rich lists'' (may be put in the supplementary materials online)}
\begin{figure}[h]
\label{fig1}
\centering
\caption{Beanplots for wealth of world billionaires, Forbes data, 1996--2012$^a$}
\includegraphics[trim = 0mm 6mm 0mm 20mm, scale=0.85]{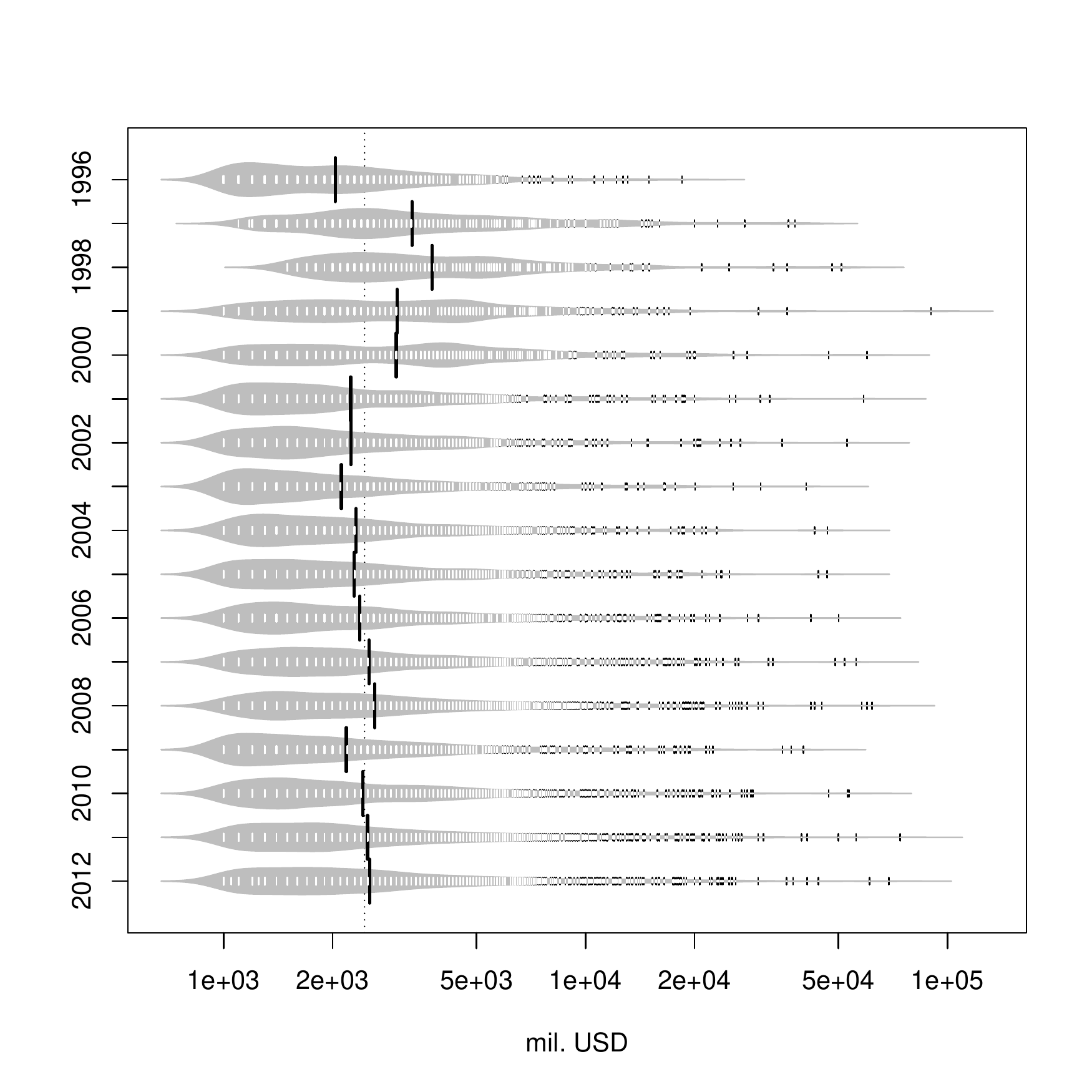}
\fnote{$^a$ The beanplot for a given year shows on a log scale individual wealth values as short vertical lines with the estimated density shown in gray. The vertical solid black lines  show mean net worth for a given year, while the overall vertical dotted line shows the grand mean.}

\end{figure}

\begin{figure}[h]
\label{fig2}
\centering
\caption{Beanplots for wealth of the US billionaires, Forbes data, 1988--2012$^a$}
\includegraphics[trim = 0mm 6mm 0mm 20mm, scale=1]{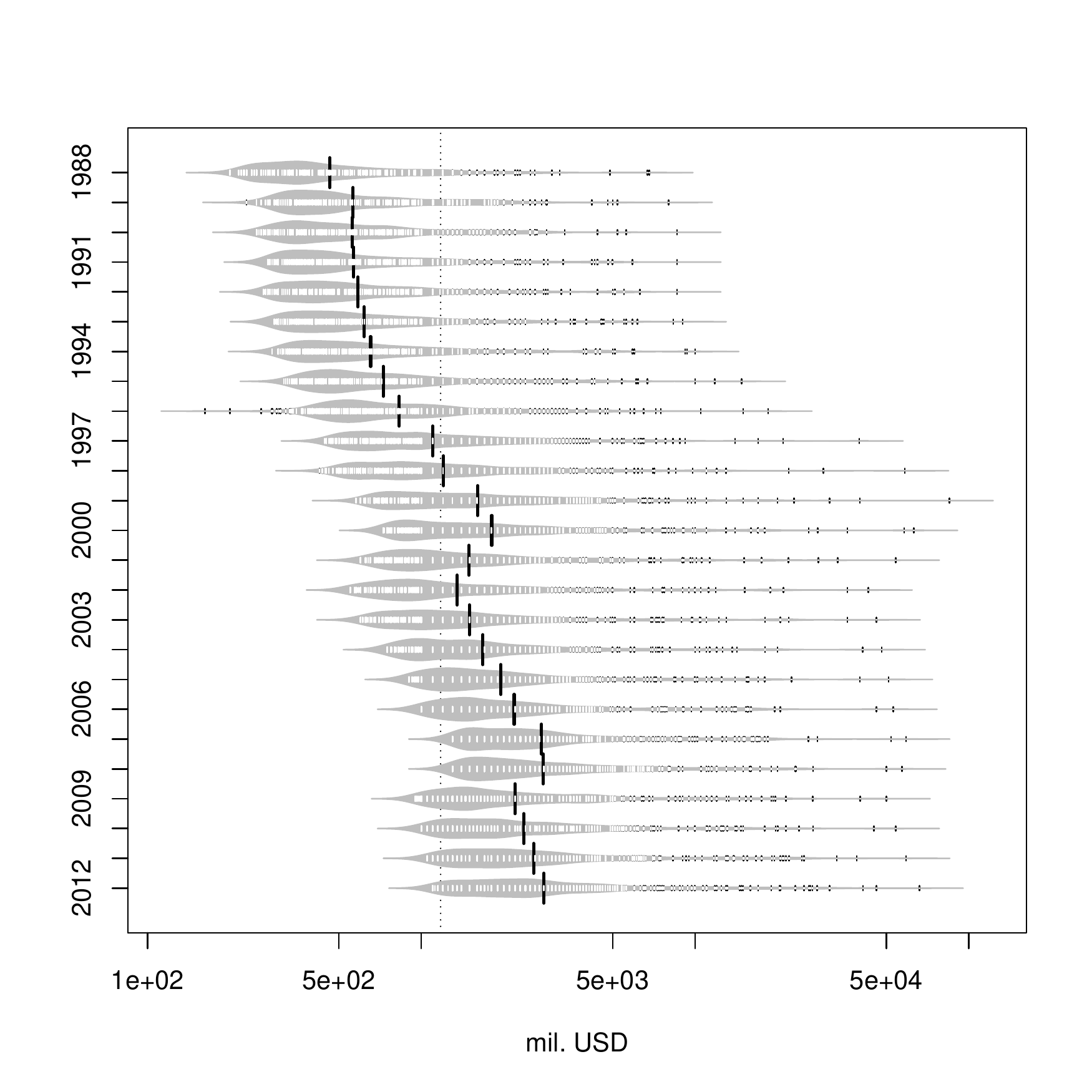}
\fnote{$^a$ See note to Fig. A.1.}

\end{figure}

\begin{figure}[h]
\label{fig3}
\centering
\caption{Beanplots for wealth of the richest Chinese, Forbes data, 2006--2012$^a$}
\includegraphics[trim = 0mm 6mm 0mm 20mm, scale=1]{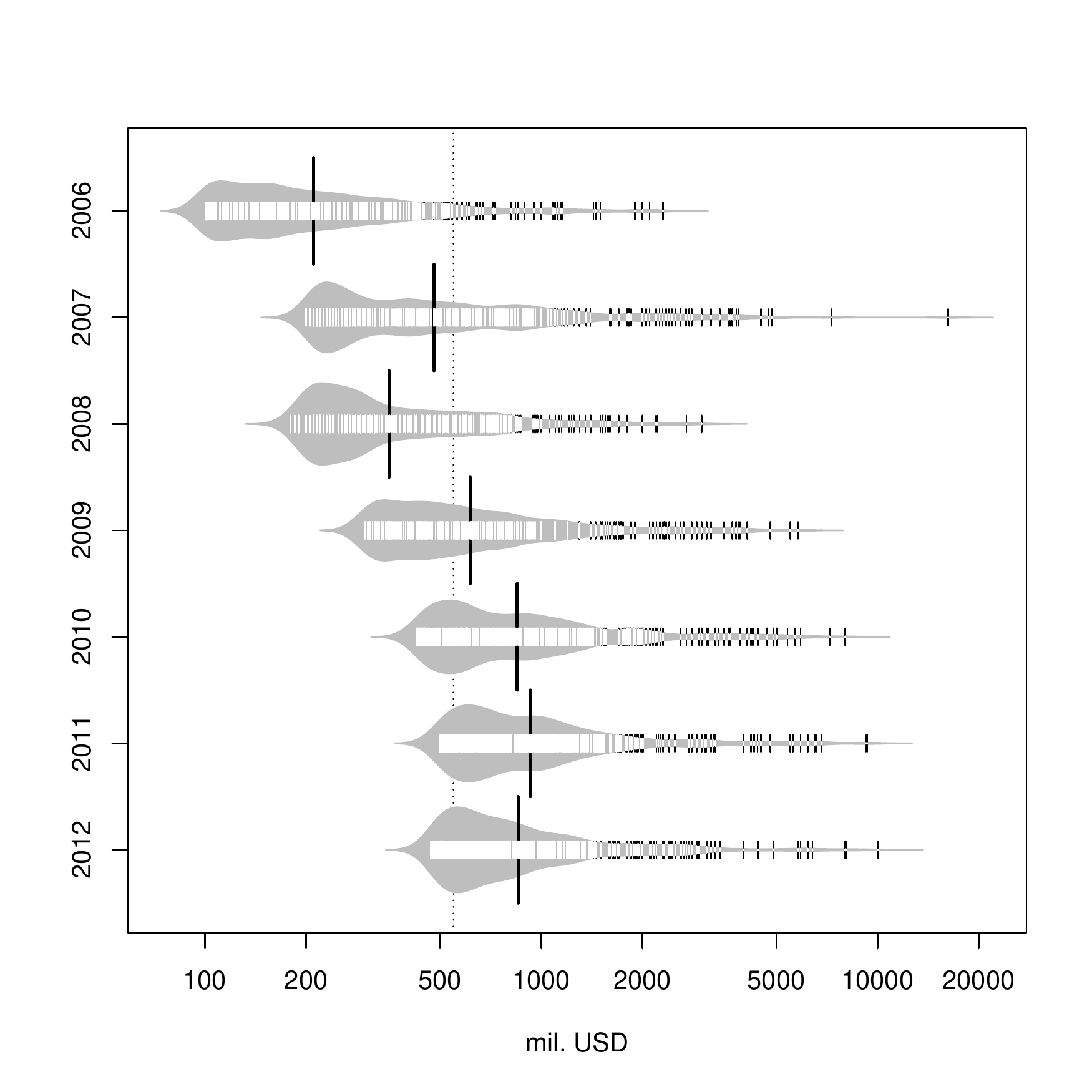}
\fnote{$^a$ See note to Fig. A.1.}

\end{figure}

\begin{figure}[h]
\label{fig4}
\centering
\caption{Beanplots for wealth of the richest Russians, Finans magazine data, 2004--2011$^a$}
\includegraphics[trim = 0mm 6mm 0mm 20mm, scale=1]{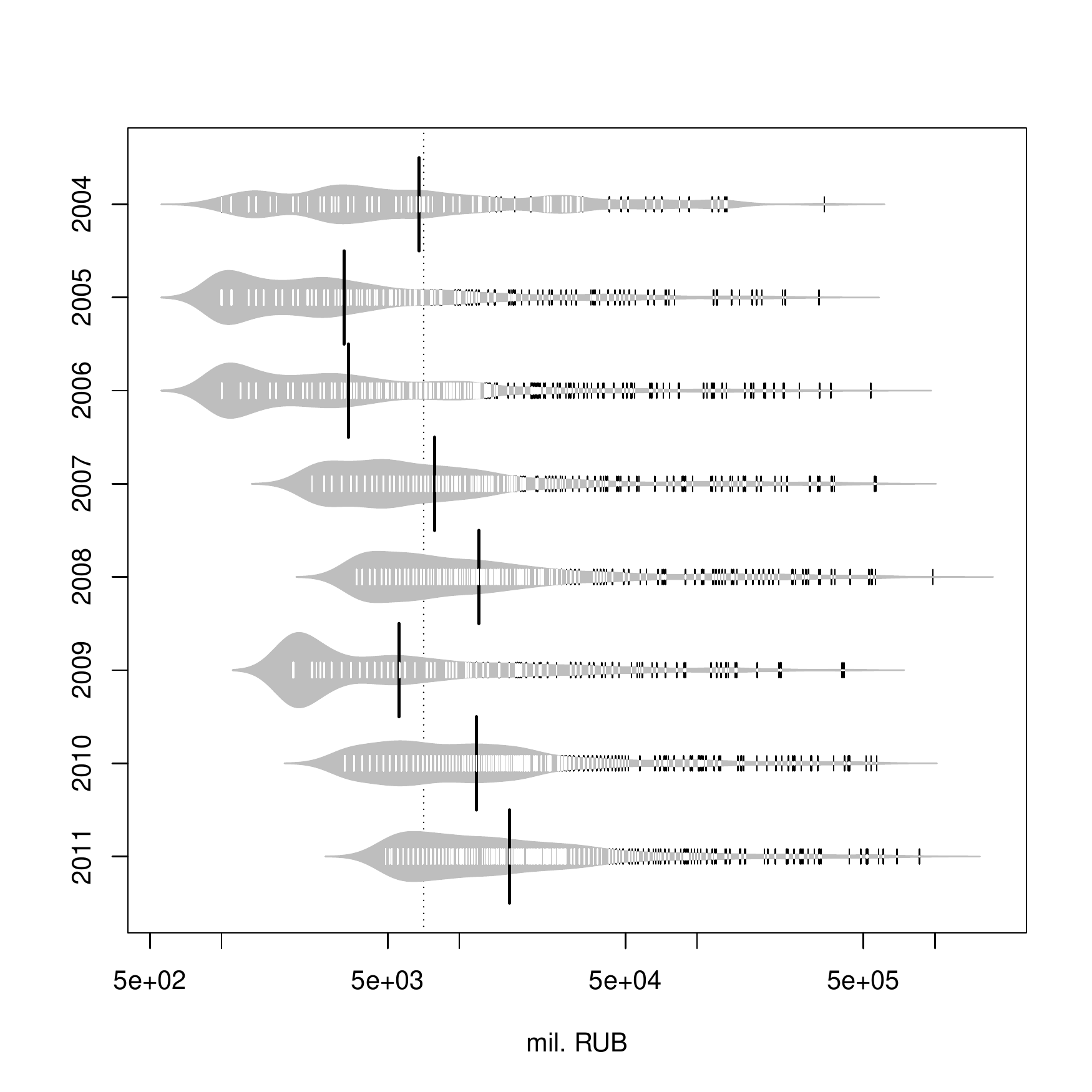}
\fnote{$^a$ see note to Fig. A.1.}

\end{figure}

\end{document}